\documentstyle[psfig,prl,aps]{revtex}

\begin{document}
\draft

\title{Spherical Shell Model  description of rotational motion}
\author{A.P. Zuker$^{*}$, J. Retamosa$^{**}$, A. Poves$^{**}$ and
   E. Caurier$^{*}$}

\address{(*)Physique Th\'eorique, B\^at 40/1 CRN,
  IN2P3-CNRS/Universit\'e Louis Pasteur BP 28, F-67037 Strasbourg
  Cedex 2, France}

\address{(**)Departamento de F\'{\i}sica Te\'{o}rica, Universidad
Aut\'{o}noma de Madrid, E-28049 Madrid, Spain}

\date{\today}

\maketitle

\begin{abstract}
  Exact diagonalizations with a realistic interaction show that
  configurations with four neutrons in a major shell and four protons
  in another ---or the same--- major shell, behave systematically as
  backbending rotors.  The dominance of the $q\cdot q$ component of
  the interaction is explained by an approximate form of SU3 symmetry.
  It is suggested that these configurations are associated with the
  onset of rotational motion in medium and heavy nuclei.
\end{abstract}

\pacs{}

\noindent
The SU3 model of Elliott \cite{EL58} provides a microscopic
description of rotors that exhibit spectra in $J(J+1)$.  For
sufficiently low $J$, or sufficiently large representations they
became perfect in the sense of having a constant intrinsic quadrupole
moment $Q_0=Q_0(J)$, where
\begin{eqnarray}
Q_0(J)={(J+1)(2J+3)\over 3K^2-J(J+1)}<JJ\vert3z^2-r^2\vert
JJ>,\eqnum{1}
\end{eqnarray}
as postulated in the strong coupling limit of the unified model of
Bohr and Mottelson \cite{BM53}.

Since the quadrupole force that appears in the SU3 Casimir operator is
also an important part of the nuclear interaction \cite{BK68,DufourZ},
we expect it to play a determinant role in the onset of rotational
motion in real nuclei - the problem we want to address.  A direct
approach would demand, in general, diagonalizations in spaces of two
major shells in neutrons and protons as first proposed by Kumar and
Baranger \cite{KUBA68}. Dimensionalities are then of order $10^{40}$,
exceeding by far what is possible at present ($10^{7}$)\cite{Caurier}.

Therefore, it is necesssary to develop a computational strategy, and
our starting point will consist in learning as much as we can from
situations in which neutrons and protons are independently restricted
to a single major shell, that can be the same close to N=Z. The exact
results in I will show that rotational features -including the
systematic appearance of backbending- are determined by the interplay
of the quadrupole force with the central field, in the subspace of a
major shell spanned by the sequence of $\Delta j=2$ orbits that comes
lowest under the spin-orbit splitting. This state of affairs is
related to the existence of an approximate symmetry (quasi-SU3),
introduced in II. The relevance of these results to rotors in medium
and heavy nuclei will be explained in III.

\begin{quote}

  {\it NOTATIONS} .  $\nu=$ neutrons, $\pi=$ protons, $C^{l
    m}=\sqrt{4\pi /(2l+1)} Y^{l m}$, $q\equiv q^{2m}=r^{2}C^{2m}$.

  $p$ is the principal quantum number, $r_p$ is the generic label for
  all orbits in the $p$-th oscillator shell {\it except} the largest
  (i.e. $j=j_{\max}=p+1/2$).

  We use $l$ for $j=l+1/2$ orbits in the sense $h=h_{11/2}$,
  $g=g_{9/2}$, $p=p_{3/2}$ etc., except in the following \\ {\it
    CONVENTION}: $pfh$ means the full $p=5$ shell, i.e.
  $p_{1/2},p_{3/2},f_{5/2,},f_{7/2},h_{9/2},h_{11/2}$, while
  $hfp=h_{11/2}f_{7/2}p_{3/2}$, and similarly for other shells.

\end{quote}

{\bf I. Exact Results.} Although a space of a full major shell, with
very specific single-particle spacings, is necessary to ensure strict
SU3 symmetry, we know of several examples where the $ds$ or $fp$
subspaces produce rotor-like spectra in the presence of spin-orbit
splittings: $(ds)^4$ describes $^{20}$Ne quite well\cite{Arima} and
$(ds)_\pi^{3,4}(fp)_\nu^2$ configurations explain the onset of
deformation in $^{31}$Na and $^{32}$Mg\cite{PovRet}. Furthermore
$^{48}$Cr provides the first example of a backbending band in N=Z
nuclei.  The experimental spectrum \cite{Cameron} is almost perfectly
reproduced by a full $(pf)^8$ shell model calculation, with strong
indications that the $(fp)^8$ space is sufficient to explain the
quadrupole coherence \cite{CZPM94}. The situation has a double
interest. As we shall explain in III, configurations that consist of 4
protons in a major shell and 4 neutrons in another (the same in
N$\approx$Z) play a key role in the onset of rotational motion in
heavier nuclei, and the restriction to the $\Delta j=2$ spaces makes
the diagonalizations possible, as illustrated by the four cases we are
going to treat (in parenthesis the corresponding m-scheme
dimensionalities):

$(fp)^8 T=0$ ; $(2\times10^4)$, \hspace{1cm} $(fp)_\pi^4(gds)_\nu^4$ ;
$(1.1\times 10^5)$,

$(gds)^8 T=0$ ; $(6 \times 10^5)$, \hspace{0.5cm}
$(gds)_\pi^4(hfp)_\nu^4$ ; $(1.9 \times 10^6$),

against

$(pf)^8 T=0$ ; $(2 \times10^6)$, \hspace{1cm} $(pf)_\pi^4(sdg)_\nu^4$
; $(10^7)$,

$(sdg)^8 T=0$ ; $(5 \times10^7)$, \hspace{0.5cm}
$(sdg)_\pi^4(pfh)_\nu^4$ ; $(1.9 \times10^8)$.

We shall compare the results obtained with the KLS
interaction\cite{KLS69} and with a pure quadrupole forces using
$\hbar\omega=9\,$MeV with a uniform single particle spacing
$\varepsilon=1\,$MeV, corresponding to the standard $-\beta\ell\cdot
s$ splitting ($\beta\approx20\rm A^{-2/3}\,MeV$ and $
\hbar\omega\approx40A^{-1/3}$)\cite{BM64}.

It has been shown in \cite{DufourZ} that for one shell the quadrupole
component of a general realistic interaction has the form $-e_2\bar
q_p\cdot\bar q_p$, where $e_2$ goes as A$^{-1/3}$ and, $\bar q
_p=q_p/{\cal N}_p^{(2)}$ is the quadrupole operator $q_p$ in shell
$p$, normalized by ${\cal N}_p^{(2)}$ - the square root of the sum of
the squares of the matrix elements of $q_p$ - which goes as
$(p+3/2)^2$.  For two contiguous shells the force is $-e_2(\bar
q_p+\bar q_{p+1})\cdot(\bar q_p+\bar q_{p+1})$, with the {\it same}
$e_2$ coupling, and it differs markedly from the traditional $\chi
(q_p+ q_{p+1})\cdot( q_p+q_{p+1})$, with $\chi=O(\rm
A^{-5/3})$\cite{BK68}.

Fig.1 shows the yrast bands in the four spaces. Rotational behaviour
is fair to excellent at low $J$.  As expected from the normalization
property of the realistic quadrupole force the moments of inertia in
the rotational region go as $(p+3/2)^2\,(p'+3/2)^2$, i.e. if we
multiply all the $E_\gamma$ values by this factor the lines become
parallel. The $Q_0$ values are constant to within 5\% up to a critical
$J$ value at which the bands backbend.

Since all the spaces behave in the same way we specialize to $(gds)^8$
in what follows. Fig.2 shows the results of diagonalizing $e_2\,\bar
q_p\cdot \bar q_p\;(p=4)$. At $e_2=9.6$ the $\varepsilon$ splittings
are overwhelmed and we have a nearly perfect rotor. The value of $Q_0$
stays practically constant up to $J=16-18$ and then decreases slowly.
At $e_2=4.8$, 3.2 and 2.4 the rotational behaviour remains very good
below $J=14$. Then there is a break and the upper values are again
aligned. At $e_2=3.2$ the overlap of each state with the one obtained
with the full KLS interaction is always better than $(0.95)^2$, which
suggests that
$$<h\vert {\cal H}\vert h>_J\approx <q\vert {\cal H}\vert q>_J,$$
where $\vert h>$ and $\vert q>$ are the eigenstates of the full
Hamiltonian ${\cal H}$ and the quadrupole force ($e_2=3.2$)
respectively. Fig.3 shows that this is the case indeed. It means that
the observed backbending pattern is obtained by doing first order
perturbation theory on $|q>$: the spectrum changes but not the {\it
  structure} (i.e. the wavefunctions).  A similar situation is found
when comparing the full $(pf)^8$ calculation with a renormalized
interaction and $\varepsilon=2$ (fig.~10 of ref.\cite{CZPM94}) and the
$(fp)^8$ result in fig.~1: the backbend occurs at the same $J$ and the
$Q_0$ values are very close in spite of a much larger slope (i.e.,
smaller moment of inertia) in the bigger calculation. Here again
perturbation theory should operate well.

To gain some insight into the backbending phenomenon we examine the
evolution of the wavefunctions and quadrupole moments. In fig.4 we
find that for $e_2$=9.6 the percentage of the $g^8$ configuration in
the full eigenstate is very small and nicely correlated with the $Q_0$
values. This is what we expect from a good rotor, for which the
amplitudes of any configuration (not only $g^8$) must be
$J$-independent (since all states must be projections of the same
intrinsic state). For the KLS results and their $e_2$=3.2 counterparts
$Q_0(J)$ decreases abruptly above $J=14$, while the $g^8$
configuration increases its amplitude and becomes dominant in the
region where $Q_0(J)$ reaches a plateau. It is clear that at the
backbend the notion of intrinsic state loses - or changes - its
meaning, and the idea of a band crossing suggested by fig.2 becomes
questionable. Work remains to be done to understand the connection of
our results with the mean field ones \cite{Egido}.

{ \bf II. Quasi-SU3.} That the build up of quadrupole coherence needs
only the lower $\Delta j=2$ sequence of the full shell can be
understood by examining table 1 where we list the matrix elements of
$q^{20}=r^2C^{20}={1\over2}(3z^2-r^2)$, in $jj$ and LS coupling. It is
seen that the $\Delta j=1$ matrix elements are small, both for $m$
small (prolate shapes) and $m$ large (oblate).  If we simply neglect
them, diagonalizing the $\Delta j=2$ matrix in $jj$ scheme is very
much equivalent to diagonalizing the exact $q^{20}$ operator in LS
scheme.  This amounts to saying that the sequence
$j=1/2,\,5/2,\,9/2\cdots$ (or $j=3/2,\,7/2,\,11/2\cdots$) must behave
very much as an $\l=0,\,2,\,4\cdots$ (or $\l=1,\,3,\,5\cdots$) one.
Therefore we introduce a new operator (the 'quasi' $q^{20}$), defined
in the $\Delta j$=2 space via the following replacements in the LS
matrix elements of $q^{20}$: $\l\to j$, $p\to p+1/2$, $m\to m+1/2$ and
$-m\to -m-1/2$: $(m>0)$.

In fig.5 we draw to the left the spectrum of the full $q^{20}$
operator (in fact 2$q^{20}$), i.e., the SU3 Nilsson orbits. The
band-heads come at $2(p+3/4-3/2\vert m\vert)$. To the right we have
plotted the spectrum of the 'quasi' 2$q^{20}$ operator. Now the
band-heads are at $2(p+1/2-3/2\vert m\vert$), that is, the exact LS
values, except for $m=\pm1/2$, where the one to one correspondence
between the ``quasi'' $q^{20}$ and the exact $q^{20}$ in LS scheme
breaks down. The corresponding ``quasi-SU3'' symmetry cannot be exact
because of this mismatch. Notice however that the mismatch is very
small ($< 1\%$). The spectrum of the true $q^{20}$ operator in the
$\Delta j=2$ space is extremely close to the one in fig.~5, and it is
clear that the amount of quadrupole coherence obtained by filling the
$m$ (or $K$) $=1/2$ and 3/2 orbits is almost as large as for the SU3
orbitals. For the eight particle blocks we are interested in, the
intrinsic $Q_0$ would be
\begin{eqnarray}
  Q_0=8[e_\pi(p_\pi-1)+e_\nu(p_\nu-1)] \eqnum{2}
\end{eqnarray}
For $p_\pi=5$, $p_\nu=6$ and effective charges $e_\pi=1.5$,
$e_\nu=0.5$, we have $Q_0=68$ (in dimensionless oscillator
coordinates, i.e., $r \rightarrow r/b$ with $b\approx A^{-1/6}fm$),
which is related to the $E2$ transition probability from the ground
state by $B(E2)\uparrow=10^{-5}{\rm A}^{2/3}Q_0^2$=1.4\, $e^2b^2$ for
$\rm A=166$. We shall see soon the relevance of this number.  The
$Q_0$ values obtained in I with the $q\cdot q$ interaction at
$e_2=9.6$ saturate the value predicted by eq.2 within $2\%$ while at
$e_2=2.4$ we still have $80\%$ of this limit.

{\bf III. The onset of rotation.}

Fig.6 proposes a schematic view of the single particle order expected
around $Z=50$, $N=82$ closure. {\it Mutatis-mutandis}, the same scheme
applies to the $Z=28$, $N=50$ and $Z=82$, $N=126$ closures. For the
lower shells we have a conventional sequence, and for the upper
(empty) ones we have assumed a spin orbit splitting, which may be
naive, but it is correct in the light nuclei and consistent with the
(scarce) available data in the heavier ones .

Nilsson diagrams \cite{Nilsson,RMP90} predict that when nuclei acquire
a stable deformation, two orbits $K=1/2$ and 3/2 -- originating in the
upper shells of fig.~5 -- become occupied. This result is common to
different regions and to different calculations and provides a strong
clue that we turn upside down when translating it into spherical
language as: {\it Rotational motion sets in when 8 particles are
  promoted to the $(pfh)_\pi^4(sdgi)_\nu^4$ configuration}, suggesting
that the calculations we have presented may be the first step in
implementing a shell model view of real rotors. The $BE2$ estimate at
the end of the previous section is a factor four smaller than the
largest observed values in the region \cite{Raman}. It means that
$Q_0$ is only a factor two too small. Therefore the lower orbits
should supply the missing half, which is very plausible because their
potential quadrupole coherence is high. The reason is that the $r_p$
groups in fig.~6 are pseudo-oscillator shells with $p'=p-1$ and the
pseudo-SU3 symmetry of Arima, Draayer, Harvey and Hecht (and hence the
left part of fig.~5) would provide the relevant coupling scheme
(ref.\cite{DR92} contains a recent survey of pseudo-SU3). It should be
noted that the quadrupole coherence in the upper(u) and lower(l)
spaces is mutually reinforced through coupling terms of the form $
-e_2\bar q_u\cdot\bar q_\l$.

{\bf Conclusions.} The zeroth order approximation we are proposing for
rotational nuclei is very similar to the weak coupling explanation of
the famous 4p-4h low lying states in $^{16}\rm O$ and $^{40}\rm Ca$.
The differences are that in the heavier nuclei they become ground
states, as in $^{80}\rm Zr$, and most probably 8p-8h excitations are
necessary to ensure the observed quadrupole coherence. The fact that
the calculations naturally produce backbending in the upper
configurations (which are the driving ones, as in light nuclei)
indicates that we are closer to real rotors than to idealized
constructions.  The tendency of quadrupole forces of Elliott type to
produce clustering in the excited states \cite{BR66,Abgrall} will
probably lead to significant differences of interpretation between the
spherical and deformed formulations for large quadrupole moments.

This work is supported in part by the IN2P3 (France)-CICYT (Spain)
agreements an by grant DGICYT PB93-263 (Spain).

\begin{table}
\caption{The matrix elements of $r^2$ and $C_{20}$ in $jj$ and LS
 coupling}
\begin{tabular}{ll}
%\tableline
$<pl\vert r^2\vert pl> =p+3/2$&$
 <pl\vert r^2\vert pl+2> =-[(p-l)(p+l+3)]^{1/2}$\\
\\
$<jm\vert C_2\vert jm> =\displaystyle{j(j+1)-3m^2\over2j(2j+2)}$&$
<l m\vert C_2\vert l m> =
\displaystyle{l(l+1)-3m^2\over(2l+3)(2l-1)}$\\
\\
$<jm\vert C_2\vert j+1m> =-
\displaystyle{3m[(j+1)^2-m^2]^{1/2}\over(2j+4)(2j+2)(2j)}$\\
\\
$<jm\vert C_2\vert j+2m> =\displaystyle{3\over2} \left\{{[(j+2)^2-m^2]
[(j+1)^2-m^2]\over(2j+2)^2(2j+4)^2}\right\}^{1/2}$&
$<l m\vert C_2\vert l +2m> =\displaystyle{3\over2}
\left\{{[(l+2)^2-m^2]
[(l+1)^2-m^2]\over(2l+5)(2l+3)^2(2l+1)}\right\}^{1/2}$\\
%\tableline
\end{tabular}
\end{table}

\begin{figure}
  \begin{center}
    \leavevmode
  \psfig{file=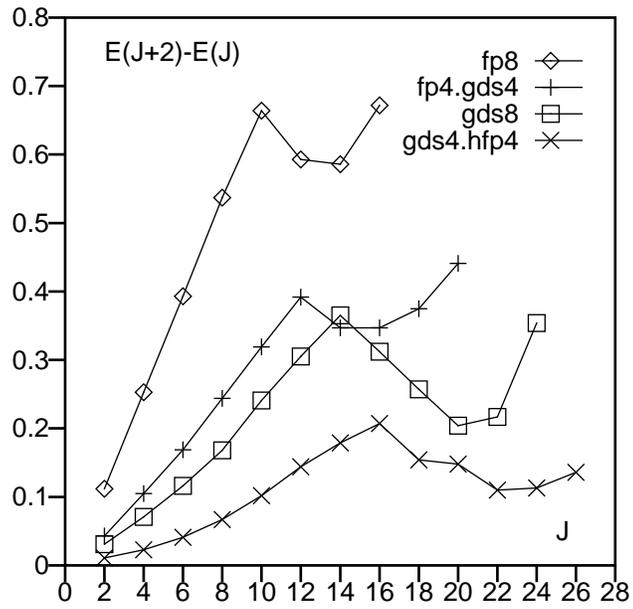,angle=270}
  \end{center}
\caption{Yrast transition energies $E_\gamma=E(J+2)-E(J)$ for different
configurations, KLS interaction.}
\end{figure}

\begin{figure}
  \begin{center}
    \leavevmode
   \psfig{file=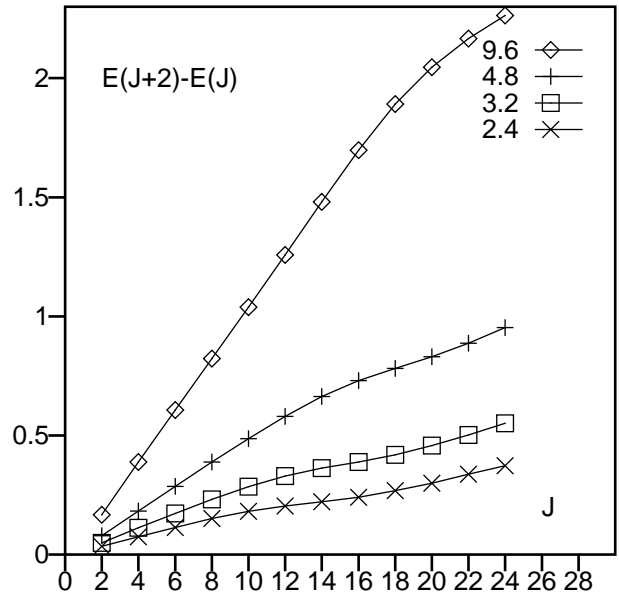,angle=270}
\caption{Yrast transition energies $E_\gamma=E(J+2)-E(J)$ for the $(gds)^8$
configuration with an $-e_2\bar q\cdot\bar q$ force.}
  \end{center}
\end{figure}

\begin{figure}
  \begin{center}
    \leavevmode
    \psfig{file=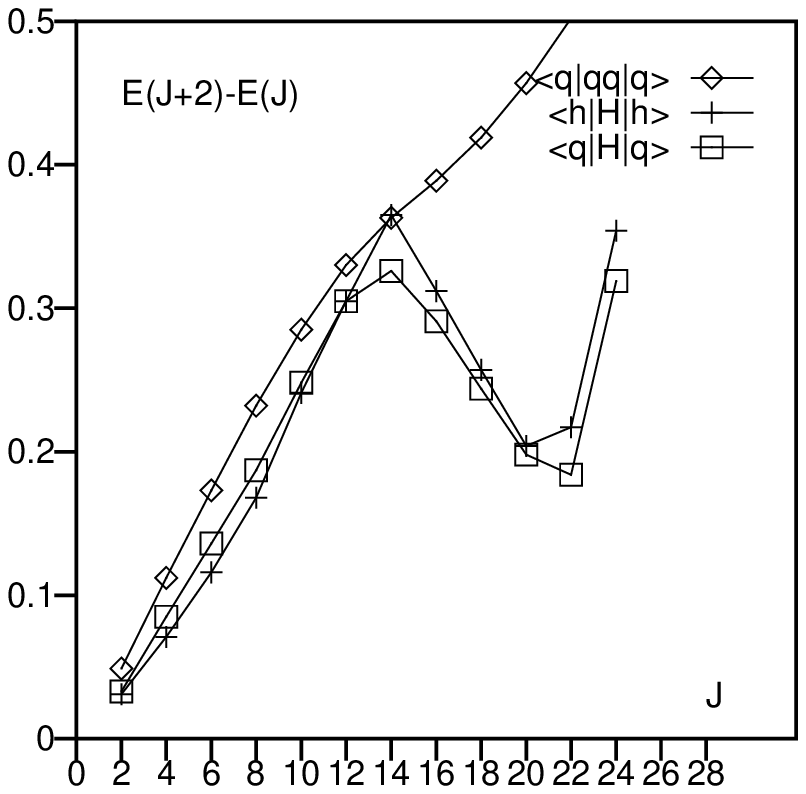,angle=270}
  \end{center}
\caption{$<h\vert{\cal H}\vert h>=(gds)^8$ in fig.~3;
          $<q\vert qq\vert q>=3.2$ in
fig.~4 compared with $<q\vert{\cal H}\vert q>$. See text.}
\end{figure}
\begin{figure}
  \begin{center}
    \leavevmode
 \psfig{file=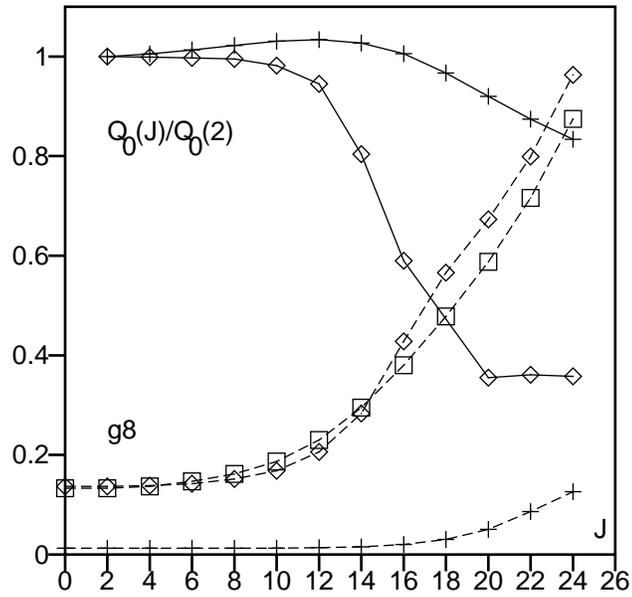,angle=270}
  \end{center}
\caption{$Q_0(J)/Q_0(2)$ (full lines) and $g_8=<g^8\vert(gds)^8>^2$
(dashed lines). Wavefunctions calculated with $-e_2\bar q\cdot\bar q$
(crosses $\equiv e_2=9.6$, squares $\equiv e_2=3.2$) and
 KLS (diamonds).}
\end{figure}
\newpage
\begin{figure}
  \begin{center}
    \leavevmode
     \psfig{file=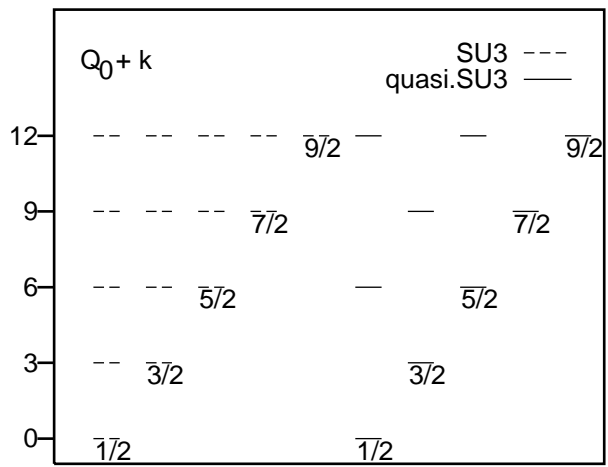,angle=270}
  \end{center}
\caption{Nilsson orbits for SU3 $(k=2p)$ and quasi-SU3
  ($k=2p-1/2$).}
\end{figure}

\begin{figure}
  \begin{center}
    \leavevmode
 \psfig{file=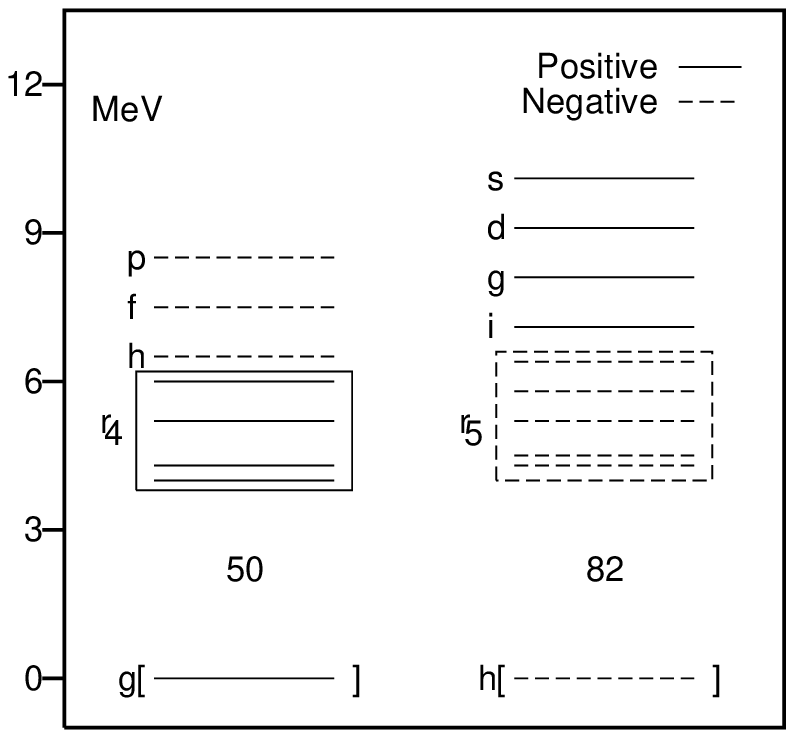,angle=270}
  \end{center}
\caption{Schematic single particle spectrum above $^{132}$Sn.}
\end{figure}

\begin{thebibliography}{99}

\bibitem{EL58} J.P. Elliott {\em Proc. Roy. Soc.} {\bf 245} 128,562
  (1956) \bibitem{BM53} A. Bohr and B. Mottelson, {\em Math. Fis.
    Medd. Dan. Vid. Selsk} {\bf 27} no 16 (1953) \bibitem{BK68} M.
  Baranger and K. Kumar {\em Nucl. Phys.} {\bf A110}, 490 (1968)

\bibitem{DufourZ} M. Dufour and A.P. Zuker, submitted to {\em Phys.
    Rev. Lett.} and preprint CRN 93-29, 1993, Strasbourg

\bibitem{KUBA68} K. Kumar and M. Baranger {\em Nucl. Phys.} {\bf
    A110}, 529 (1968) \bibitem{Caurier} ANTOINE code, CRN, Strasbourg
  1989, to be released \bibitem{Arima} A. Arima, S. Cohen, R.D. Lawson
  and M. McFarlane {\em Nucl. Phys.} {\bf A108} 94 (1968)

\bibitem{PovRet} A. Poves and J. Retamosa {\em Nucl. Phys.} {\bf
    A571}, 221 (1994)

\bibitem{Cameron} J.A. Cameron {\it et al.}, {\em Phys. Lett.} {\bf
    319B} 58 (1993) \bibitem{ACZ91} A.~Abzouzi, E. Caurier and A.P.
  Zuker {\em Phys. Rev. Lett.} {\bf66}, 1134 (1991) , and in
  preparation.

\bibitem{CZPM94} E. Caurier, A.P. Zuker, A. Poves and G.
  Martinez-Pinedo, {\em Phys. Rev.} {\bf C50}, 225 (1994)

\bibitem{KLS69} S. Kahana, H.C. Lee and C.K. Scott {\em Phys. Rev.}
  {\bf 180}, 956 (1969) ; {\em Phys. Rev.} {\bf 185}, 1378 (1969)
\bibitem{BM64} A. Bohr and B. Mottelson {\em } Nuclear Structure
  vols~I (Benjamin, 1964)

\bibitem{Egido} J. L. Egido and L. M. Robledo, Phys. Rev. Lett. {\bf
    70}, 2876 (1993)

\bibitem{Nilsson} S.G. Nilsson {\em Mat. Fys. Medd. Dan. Vid. Selsk.}
  {\bf 29}, 1 (1955) \bibitem{RMP90} A.K. Jain, R.K. Sheline, P.C.
  Wood and K. Jain {\em Rev. Mod. Phys.} {\bf 62} 393 (1990)

\bibitem{Raman} S. Raman, C.W. Nestor and K.H. Bhatt {\em Phys. Rev.}
  {\bf C37}, 805 (1988)

\bibitem{DR92} J.P. Draayer in Future Directions in Nuclear Physics
  with $4\pi\gamma$ Detectors (J.Dudek and B. Haas eds A.I.P. 1992)

\bibitem{BR66} D.M. Brink in Course XXXVI Enrico Fermi School,
  C.~Bloch editor (Academic Press, 1966) \bibitem{Abgrall} Y. Abgrall,
  G. Baron, E. Caurier and G. Monsonego, Phys. Lett. {\bf 24B}, 609
  (1967); Phys. Lett. {\bf 26B}, 53 (1967)
\end{thebibliography}
\end{document}